\documentclass[fleqn,11pt,a4paper]{article}
\usepackage{epsfig}

\newcommand{\lsim}{\rlap{\raise 2pt \hbox{$<$}}{\lower 2pt \hbox{$\sim$}}}
\newcommand{\gsim}{\rlap{\raise 2pt \hbox{$>$}}{\lower 2pt \hbox{$\sim$}}}
\newcommand{\etal}{{\it et al.}}
\newcommand{\ie}{{\it i.e.\ }}
\renewcommand{\d}{\mbox{\rm d}}
\newcommand{\nmu}{{N_{\mu}}}
\newcommand{\lt}{{\Lambda_T}}

\newcommand{\lca}{{\Lambda_{cal}}}
\newcommand{\labs}{{\Lambda_{abs}}}

\newcommand{\ffig}[5]{\begin{figure}[#1]\vfill\begin{center}
\mbox{\epsfig{figure=#2,width=#3}}\caption{#4}\label{#5}
\end{center}\vfill\end{figure}}

\begin{document}
\noindent
\noindent
TSL/ISV-95-0138   \hfill ISSN 0284-2769\\
April 1996      \\ 
\\
\vspace*{5mm}
\begin{center}
  \begin{Large}
  \begin{bf}
Searching for Heavy Quarks at NMC:\\
A Simple Estimate of Background \\
Muons from $\pi$ and $K$ decays\\
  \end{bf}
  \end{Large}
  \vspace{5mm}
  \begin{large}
 M. Thunman\footnote{thunman@tsl.uu.se} \\
  \end{large}
  \vspace{3mm}
Dept. of Radiation Sciences, Uppsala University,
Box 535, S-751 21 Uppsala, Sweden\\
  \vspace{5mm}
\end{center}
\begin{quotation}
\noindent
{\bf Abstract:}
New physics can be searched for in deep inelastic scattering
experiments. New phenomena, like additional production mechanisms for
heavy quarks, manifest themselves through production and decays of
short-lived hadrons, which give additional muons in their decays. Muons
from decays of light mesons can then be a severe background, especially
at fixed target experiments. In order to make a first estimate of such
a background a simple analytic method is here developed and applied on
the muon scattering experiment NMC.
\end{quotation}

\section{Introduction}

Fixed target deep inelastic scattering (DIS) experiments like EMC/NMC
and E665 can be used for searches of new physics, even though they are
constructed for precise measurements of the structure functions. An
example of such new physics is intrinsic charm \cite{Brodsky80}, which
have been searched for by European Muon Collaboration (EMC). They based
their study on di- and tri-muon events and found an upper limit of
0.6\% for the intrinsic charm probability \cite{IC-emc}, \ie the
probability that the proton wave-function is in the intrinsic charm
state $|uudc\bar{c}\rangle$. The data have been reanalysed and the
newer analyses confirms the conclusion of the original analyses
\cite{Hoffman,Harris}.

New physics like intrinsic charm and bottom would manifest themselves
through the production of heavy hadrons containing these quarks. The
heavy hadrons would then decay giving rise to additional 
muons. To extract such a signal, other sources
of muons must be well understood. These sources are 
decays of hadrons containing perturbatively 
produced heavy quarks, \ie by perturbative QCD (pQCD),
and decays of light mesons. This background has to be well
understood and not too large if it should be possible to extract a
signal. The perturbative production of
heavy quarks is reasonable well understood, and its understanding could
in fact be improved by a study of heavy quark production. The
background from decays of light mesons is, besides the production
mechanism, also dependent on the lay-out of the experiment. Light
mesons have rather long decay lengths and could therefore be absorbed
in various material before they decay. A simple analytic first estimate
of the absorption effects is therefore desirable, before a full
simulation of this source of muons is performed.

There are in principle two methods to reduce the background from decays
of pions and kaons. Either one can build the experiments so small that
the probability that meson decays between the interaction point and the
detector become small enough, or one uses a larger set up with
hadron absorbers to absorb the light mesons. Both EMC and NMC (New Muon
Collaboration) use rather large experimental set-ups because they
are designed for precise measurements of the muon momentum. Therefore,
only the later method can be used.

In the mentioned intrinsic charm search at EMC \cite{IC-emc} a long
($\sim 3.75\,m$) Fe-scintillator calorimeter was used as target
\cite{emcdimuon}. The first three meters of the calorimeter was used as
target for the study, and the remaining part was used as a
calorimeter/absorber ($\sim 5$ absorption lengths).  This target was
not used by NMC, instead different homogeneous targets of hydrogen,
deuterium, carbon and different metals were used.  Because of the high
interaction rate in the heavier targets the number of tracks in the
tracking region was expected to be unnecessary high. The reduction of
the number of tracks, both from mesons and from muons coming from light
meson decays, was done by putting a calorimeter close to the beam axis
in the target region and a calorimeter/absorber directly after the last
part of the target (see Fig.\,\ref{fig:experi}). This arrangement,
which was necessary for the original purpose of the experiment, also
makes it useful for a search of heavy quark production in terms of
additional muons from their decays.

Deep inelastic electron-nucleon scattering experiments
are not as feasible as muon scattering experiments for searching
of intrinsic charm. This because the scattered electron would also be
absorbed together with the light mesons in the calorimeters/absorbers.
Cuts in variables related to the scattered lepton are valuable to
distinguish intrinsic charm events from other events, which is not
possible if the electron is lost. However, by only measuring the muon
rate and their energy spectrum valuable indications for intrinsic charm
can still be found.

The lay-out of the paper is the following. First some general aspects
of meson and muon production are discussed (section~2). Then the
experimental set-up of the NMC experiment is described (section~3) and
the corresponding muon flux from light meson decays is calculated
(section~4). The possibility to use cuts in parameters  related to
secondary vertices to further reduce the background is discussed in
section~5. The analysis is concluded (section~6) with some remarks.

\section{General aspects of meson and muon production in DIS}

The starting point of the analysis is a calculation of the muon rate
from light meson decays, where attenuation of mesons in absorbing
material in the experimental set-up is neglected. This calculation
could either be done with a Monte Carlo, as in this study, or with an
approximate analytical method. That the calculation is performed
without absorption of mesons means that the mesons are allowed to decay
in a distance of the size of the experiment. For a fixed target deep
inelastic scattering experiment like NMC this distance  is $\sim
20\,m$. The goal of the analysis is to calculate a factor which
the originally calculated flux should be multiplied with in order to
include absorption of mesons. This suppression factor ($\xi$) is the
fraction of muons reaching the detectors in a set-up with absorbing
material compared to without.

If the size of the experimental set-up is small compared to the decay
length of the pions and kaons, decays can be neglected when calculating
the development of the  pion and kaon flux. Since muons of energy lower
than $\sim10\,GeV$ cannot be identified in NMC, meson energies lower
than $\sim15\,GeV$ are not of interest. At such low energies $\sim3\%$
of the $\pi$ and $\sim15\%$ of the $K^{\pm}/K^0_L$ will decay in a
length of $20\,m\ (L_{tot})$. This means that attenuation due to decays
is negligible compared to the total flux, which therefore is
approximately constant within the `empty apparatus'. The above aimed
method, based on a suppression factor, is therefore applicable. Since
the light meson fluxes are roughly constant within the empty parts of
the  apparatus, the muon production rate is also constant. A muon
production rate per meter, $\kappa$, is now defined by dividing the
muon flux from the Monte Carlo with the length scale ($L_{tot}$), \ie
the number of muons produced per meter in an `average Monte Carlo
event'.

Muon nucleon scattering is dominated by soft processes ($Q^2\approx0$),
\ie Coulomb scattering. Since typically only small amounts of energy is
exchanged in such an interaction, the muon is only deflected a small
angle with small energy loss. This means that muon fluxes are not
attenuated by such interactions. In deep inelastic muon proton
interactions, on the other hand, the momentum transfer is large, and
the muon is scattered at a larger angle with sizeable energy loss. The
cross section for this type of interactions is small, which implies
that it does not cause any strong attenuation of the muon beam. So
attenuation of beam, scattered and secondary (from decays) muons due to
interactions can be safely neglected. Since the attenuation of the
muon beam is neglected, the muon interaction rate and thereby the meson
production rate is constant along the target (and along the
calorimeter/absorber). A beam muon interaction rate is now defined,
$\zeta=1/L$, were $L$ is the target length ($\zeta$ is normalised to
one event per target). The meson production rate, $\eta$, is now
obtained by multiplying $\zeta$ by the number of produced mesons in a
typical deep inelastic scattering event, which is unity if $\eta$
is given in units of `average Monte Carlo event'.

The pions and kaons produced in the primary interaction will interact
in the target and the meson flux will therefore be attenuated when
passing through the target. The meson flux will in the same way be
attenuated in the calorimeter and the absorber material. The inelastic
pion-proton and kaon-proton interaction cross sections are roughly the
same and energy independent over the relevant energy range
($15-100\,GeV$) and have the value $\sigma\approx20\,mb$ \cite{PDB}. A
material independent interaction thickness is given by
\begin{equation} 
\lambda = \frac{\rho}{\sigma\ n_N} = \frac{1}{\sigma\ N_A} = 83 g/cm^2, 
\end{equation}  
where $\rho$ is the density, $n_N$ is the number density of nucleons
and $N_A$ is the Avogadro number. Any cross section dependence on the
target number is assumed to be of the form $A^1$, where $A$ is the
target mass number. A material dependent interaction length is now
defined 
\begin{equation}  
\Lambda = \frac{\lambda}{\rho}.  
\end{equation}

The meson flux develops in the target according to
\begin{equation}
\frac{\d \phi(z)}{\d z}\ =\ \eta\, -\, \frac{1}{\lt}\, \phi(z),
\end{equation}
where $z$ is the coordinate along the beam axes.
$\eta$ is the production term and $-\phi(z)/\lt$ is the
attenuation due to interactions in the target ($T$). The solution is
\begin{equation}
\phi(z)\ =\ \eta\, \lt\, (1-e^{-z/\lt}),
\label{eq:piflux}
\end{equation}
where no initial meson flux is assumed. $\phi$ is here given in units
of `average Monte Carlo event'.
The contribution to the muon
flux due to meson decays in the target is given by
\begin{equation}
\nmu\ =\ \int^{L}_{0} \d z\ \kappa\, \phi(z)\ =\ \kappa\,\eta\, \lt\,
(L-\lt\,(1-e^{-L/\lt})),
\end{equation}
where $L$ is the length of the target. So far it has been assumed that
no particles escapes from the target through its lateral surface. This
is an approximation that is going to be discussed further when the
experimental set-up is specified.

In a region between a target and a calorimeter/absorber the meson flux
is  unaffected by absorption, but it contributes to the muon production
through the pion and kaon decays. The decays give rise to the muon
flux
\begin{equation}
\nmu\ =\ \kappa\,\ell\,\phi_0,
\end{equation}
where $\ell$ is the length between the target and the 
calorimeter/absorber.

When the meson passes through a calorimeter or an absorber the flux is
attenuated due to interactions. The meson flux develops according to
\begin{equation}
\phi(z)\ =\ \phi_0\, e^{-z/\Lambda},
\end{equation}
where $z$ is the depth in the calorimeter/absorber.
Mesons will also decay in the calorimeter/absorber and therefore also
contribute to the muon flux by
\begin{equation}
\nmu\ =\ \kappa\ \phi_0\ \Lambda\ (1-e^{-\ell/\Lambda}),
\end{equation}
where $\ell$ is the length of the calorimeter/absorber.

\section{The experimental set-up} 

The target set-up that is going to be used for the calculation is the
heavy target one used by NMC in 1988. The elements relevant for the
muon production will be specified whereas other elements will only be
briefly discussed.

The target system consists of two complementary set of targets which
are interchangeable. Each set consist of four targets
(Target\,1--Target\,4 in Fig.\,\ref{fig:experi}, only one set shown).
The target material is either homogeneous carbon or evenly distributed
slices of different metals.  The effective thickness of the different
targets is roughly the same ($\sim145\,g/cm^2$). Carbon is used as
reference and two targets are always made of homogeneous carbon,
target~1 and 3 or 2 and 4 in the two sets of targets, respectively. The
calculation will be performed for a set-up where all targets are made
of a homogeneous material, \ie all targets made of carbon which
simplifies the calculation. This is a reasonable approximation since
the targets were designed to be as equivalent as possible. The lengths
of each target and the distances between them are shown in Fig.\,
\ref{fig:experi}.

\ffig{b}{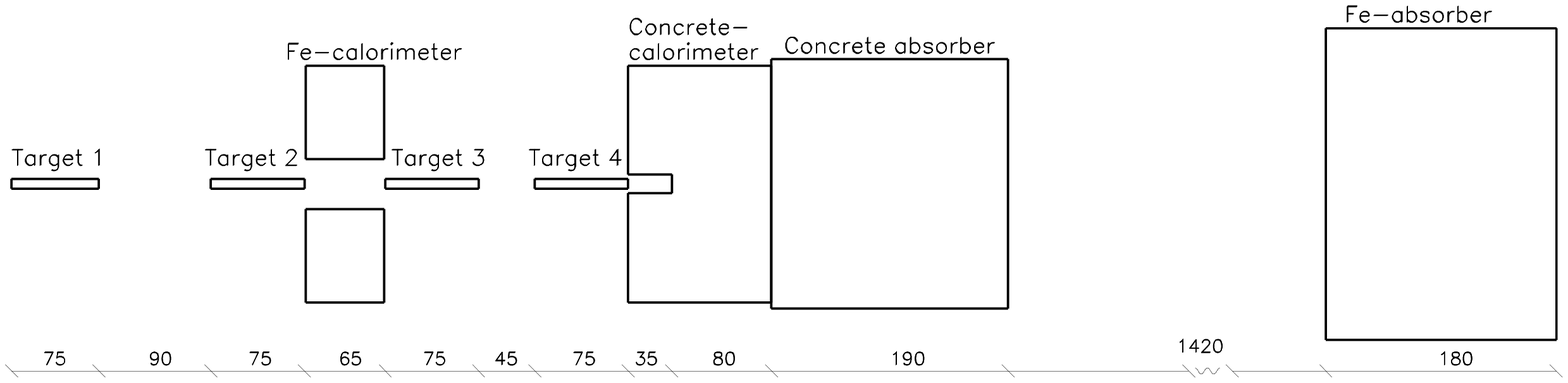}{15cm}
{\it Lay out of the relevant parts of the experiment. All lengths are
in cm.} {fig:experi}

Between Target~2 and 3 an iron calorimeter is placed. It consist of layers
of iron and plastic scintillator with a central hole for the beam
($52*26\,cm^2$) and has a total thickness of $325\,g/cm^2$. Its purpose
is to measure the energy of hadrons coming from interactions in
Target~1 and 2.  Similarly, a calorimeter consisting of layers of
concrete and plastic scintillator is placed after Target~4.  It has a
thickness of $235\,g/cm^2$, with the exception of the central part
which only has a thickness of $157\,g/cm^2$. The reason for the lower
thickness in the central part is that a hole ($8*8\,cm^2$) is made in
the calorimeter (see  Fig.\,\ref{fig:experi}) in order to make it
possible to separate muons scattered in Target\,4 from those scattered
in the calorimeter.  A concrete absorber of thickness  $460\,g/cm^2$ is
placed after the calorimeter. Its purpose is to absorb hadrons in order
to minimize the number of particles in the downstream muon detectors.
Further downstream an iron-absorber  ($\sim400\,g/cm^2$) is placed. Its
purpose is to absorb hadrons in order to ensure that only muons reach
the detector planes after it.

The main parts of the experiment, besides the targets, are the
analyzing magnet and a number of thin detector planes for position
measurements. These elements are however not important for this
calculation and therefore not included in Fig.\,\ref{fig:experi}. The
analyzing magnet is positioned $\sim1\,m$ after the concrete absorber.
The detector planes are in three groups. The first group consists of
planes before, inside and directly after the analysing magnet. The
purpose of this group is to reconstruct tracks through the magnet. The
second group is a set of planes directly before the iron absorber,
whose purpose is, together with the first group, to measure the muon
momentum with high precision. The third group is placed after the iron
absorber and is used for muon identification.

\section{Calculation of muon fluxes}

The starting point of the analysis is a calculation of the muon
flux in which absorption effects are neglected. This has been done with
the Monte Carlo {\sc Lepto\,6.4} \cite{Lepto} which simulates deep
inelastic lepton-nucleon scattering. It uses leading order matrix
elements and parton showers for the perturbative QCD calculations of the
partonic state, and the Lund string model \cite{Lund} for the
non-perturbative hadronisation of the partons. The simulation has been
done with a $280\,GeV\ \mu^+$ beam, which is the condition under which
NMC has taken most of their data. Two sets of simulations have been
done, one with a minimal set of experimental cuts for NMC
($Q^2>2\,GeV^2, W^2>100\,GeV^2, 0.2<y<0.8$ and $E_{\mu'}>10\,GeV$), and
one with an additional set of cuts for searching for intrinsic charm
($x>0.2$ and $y<0.65$). 

To investigate if parameters related to the production vertices, in
particular the impact parameter, could be used for further
discrimination, also muons from decays of charmed particles have been
studied. Both the normal pQCD and the hypothetical intrinsic charmed
production mechanisms are considered as sources of charmed particles.
The intrinsic charm study has been performed with {\sc Lepto}, while
the pQCD study has been performed with {\sc Aroma} \cite{Aroma}. {\sc
Aroma} is a program based on {\sc Lepto}, but with an improved heavy
quark treatment due to explicit inclusion of the heavy quark masses in
the matrix element.

The muon flux gets contributions from interactions in all targets and
also from interactions in the concrete calorimeter and the concrete
absorber. It is assumed that it is possible to determine from which
target the scattered muon originated, and that the targets therefore
can be treated treated as independent. This is important if some of the
targets must be excluded from an experimental analysis because of too
high background. As mentioned earlier the interaction rate is
normalised to one event per target in order to be able to compare
directly with the Monte Carlo.

The situation is similar for all four targets. They all give
contributions to the muon flux from decays within the target, decays in
an empty section after the target and decays in a calorimeter after the
empty section. After a few simplifications the muon fluxes from them
can therefore be calculated with the same formula. The simplifications
made are that all mesons are assumed to miss the next target and are
allowed to travel freely until they hit a calorimeter. The angle a
mesom most make with the beam axis in order to miss the next taget is
$\sim3^{\circ}$. From Fig.\,\ref{fig:theta} one can see that many
mesons make so small angle with the beam axis, especillay with the
minimal set of cuts, that they hit the next target. Therefore, the 
contributions from decays in the empty sections will be  overestimated,
by a factor maybe as large as 3. Secondly it is assumed that all mesons
from Target~1 hit the Fe-calorimeter. Since the hole in the
Fe-calorimeter is rather large, this will underestimate the muon flux
in case of Target~1. It is also assumed that absorption in the
Fe-calorimeter of mesons originating from  Target\,2  can be neglected
and analogously that absorption in the first part of the concrete
calorimeter of mesons originating from Target\,4 can be neglected.
Finally, the contributions to the muon flux from decays after the
Fe-calorimeter in case of Target~1 and from decays after the concrete
calorimeter in case of Target\,2 and 3 are neglected. In case of
Target\,4, the calorimeter part is  not thick enough to neglect decays
after it, so decays in the concrete absorber is also  taken into
account, but decays after the absorber is neglected. These last
simplifications will underestimate the muon flux with a few percent.
Since the goal of this study is only to estimate the suppression due to
absorption, the precision given by these simplifications is sufficient.

\ffig{tb}{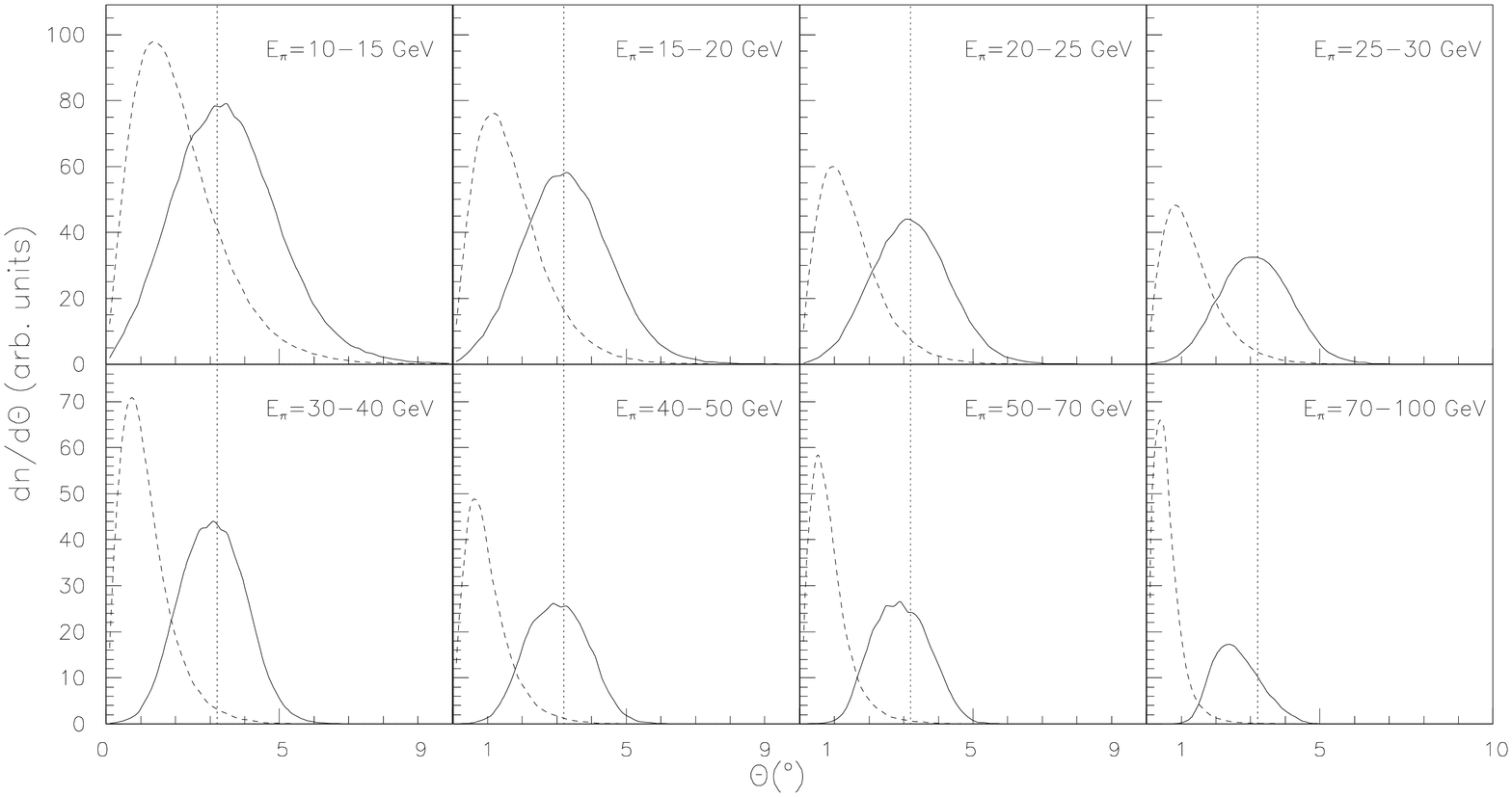}{14cm}
{\it Angular distribution of pions for different pion energy bins.
Shown are both the distribution for the minimal set of cuts at NMC
(dashed lines) and for the set of cuts applicable for a search of
intrinsic charm (solid lines). The simulations are for a 280\,GeV muon
beam and the distributions are given with an arbitrary normalisation,
which is the same within each of the two sets. Marked is also the
minimum angle $(3.06^{\circ})$ a meson must must make with the beam
axis in order to be able to escape through the lateral
surface.}{fig:theta}

For the derivation of the formulas in section~2 it was assumed that no
particles escape through the lateral surface of the targets. The
approximation is valid for a wide target and will be shown to be
reasonable in the case of NMC. The targets are cylinders of length
$L=75\,cm$ and radius $R=4.0\,cm$. A particle produced in the beginning
of the target must make an angle angle larger than $3.06^{\circ}$ with
the beam axis in order to escape through the lateral surface. In
Fig.\,\ref{fig:theta} the angular distributions of pions for different
energy bins for the two sets of experimental cuts are shown as obtained
from Monte Carlo simulations. One can see that most pions in the
case of minimal cuts fulfill this criteria. For the intrinsic charm search
most pions make an angle less than $6^{\circ}$ with the beam axis,
which means that only pions produced in the first half of the target
can escape through the lateral surface. These pions must travel through
larger amounts of material than an average pion produced more downstream
and escaping through the end-surface. The former will therefore suffer
larger absorption and therefore be less important. This, together with
the fact that still half of the pions makes an angle less than
$3^{\circ}$ with the beam axis, justify the approximation even in the
intrinsic charm search case.

The muon flux is now obtained by calculating the contributions from the
different sections of the experiment and adding them.
The contribution from meson decay within the target is 
\begin{equation}
N_{\mu,1}\ =\ \kappa\, \eta\, \lt\, (L-\lt (1-e^{-L/\lt})),
\end{equation}
where all variables have been defined earlier.
The meson flux after the target is given by Eq.\,(\ref{eq:piflux}) with
$z=L$,
\begin{equation}
\phi_M(L)\ =\ \eta\, \lt\, (1-e^{-L/\lt}).
\end{equation}
In the empty section after the target it give rise to the contribution
\begin{equation}
N_{\mu,2}\ =\ \kappa\, \eta\, \lt\, \ell\, (1-e^{-L/\lt}),
\end{equation}
where $\ell$ is the length of the empty section. Since attenuation due
to decays is neglected when calculating the evolution of the flux, the
same meson flux will reach the calorimeter. The contribution to the
muon flux from decays in the calorimeter is 
\begin{equation}
N_{\mu,3}\ =\ \kappa\, \eta\, \lt\, \lca\, (1-e^{-L/\lt})
(1-e^{-\ell_{cal}/\lca}),
\end{equation}
where $\ell_{cal}$ is the length of the calorimeter. For Target~4 the
contribution from decays in the absorber must also be included, which
is
\begin{equation}
N_{\mu,4}\ =\ \phi_0\, \kappa\, \labs\, (1-e^{-\ell_{abs}/\labs}),
\end{equation}
where
\begin{equation}
\phi_0\, =\ \eta\, \lt\, e^{-\ell_{cal}/\lca}\, (1-e^{-L/\lt})
\end{equation}
is the flux at the boundary of the calorimeter and the absorber. 

The relevant data for the experimental set-up are collected in
Table\,\ref{tab:target}. By adding the four contributing sources the
total muon flux is obtained, \ie
\begin{equation}
N_{\mu}\ =\ \sum_{i=1}^4\ N_{\mu,i}\,.
\end{equation}
Since the muon production rate $\kappa$ is normalised with respect to
the `average Monte Carlo event' (MC-flux divided by $L_{tot}$) and
$\zeta$ is normalised to one event per target this is the suppression
factor, \ie $\xi=\nmu$. The multiplicative
suppression factors for the four targets are given in
Table\,\ref{tab:target}.

\begin{table}
\begin{center}
\begin{tabular}{|l|cc|c|cc|cc|c|}
\hline
 & \multicolumn{2}{c|}{Target}   &
\multicolumn{1}{c|}{Empty section} & \multicolumn{2}{c|}{Calorimeter} & 
\multicolumn{2}{c|}{Absorber}  & \multicolumn{1}{c|}{$\xi$} \\   
 & $L$ & $\lt$ & $\ell$ & $\ell_{cal}$ & $\lca$ & $\ell_{abs}$ &$\labs$
 & \\ 
 & \multicolumn{2}{c|}{(cm)}  &
\multicolumn{1}{c|}{(cm)} & \multicolumn{2}{c|}{(cm)} & 
\multicolumn{2}{c|}{(cm)}  & \multicolumn{1}{c|}{} \\  
\hline
Target 1 & 75 & 43 & 165 & 55  & 21 &     &    & 0.055 \\
Target 2 & 75 & 43 & 260 & 115 & 55 &     &    & 0.084 \\
Target 3 & 75 & 43 & 120 & 115 & 55 &     &    & 0.051 \\
Target 4 & 75 & 43 &  35 & 80  & 55 & 190 & 41 & 0.033 \\
\hline
\end{tabular}
\end{center}
\caption{\em Data on targets, calorimeters and absorber and the
suppression factor for the final flux for the four different targets.}
\label{tab:target}
\end{table}

Analogous to the production within the targets, the primary muon can 
interact with the material in the concrete calorimeter and the concrete
absorber. The analysis is similar to the target analysis but decays in
the section between the concrete and the Fe-absorber must be included.
For the calorimeter the flux is
\begin{eqnarray}
\lefteqn{\kappa\, \eta\, \lca\, \left( \ell_{cal}-\lca (1-e^
{-\ell_{cal}/\lca})\ +\ \lca (1-e^{-\ell_{cal}/\lca}) (1-e^
{-\ell_{abs}/\labs})\ + \right.} & 
\begin{array}{c} \ \\ \ \end{array} & \nonumber  \\
\lefteqn{\left. \ell e^{-\ell_{abs}/\labs} (1-e^{-\ell{cal}/\lca})
\right), } &
\begin{array}{c} \ \\ \ \end{array} &
\end{eqnarray}
where the three different parts are from the decays in the calorimeter,
decays in the concrete absorber and decays between the absorbers
respectively. $\ell_{cal}$ is the part of the calorimeter that is in
the beam ($80\,cm$) and works as a target, and $\ell$ is the length
between the concrete  and the Fe-absorber.  From interactions in the
concrete absorber one gets the flux
\begin{equation}
\kappa\, \eta\, \labs \left( \ell_{abs}-\labs
(1-e^{-\ell_{abs}/\labs})\ +\  \ell (1-e^{-\ell_{abs}/\labs}) \right),
\begin{array}{c} \ \\ \ \end{array}
\end{equation}
where the first part is from decays within the concrete absorber and
the later part is from decays between the absorbers. In both these two
cases $\eta$ is redefined compared to the target case in order to be
normalised to one interaction in the `target'. The suppression factor
for the calorimeter is 0.027 and for the absorber 0.17.

\section{Impact parameter signature for secondary vertices}

Since the difference in life-time between charmed particles and light
mesons is large, they will typically decay at very different distances
from the primary vertex. Since the heavy hadrons are very short-lived
the muons from their decays will originate almost from the primary
vertex ($\lsim1\,cm$). Light mesons, on the other hand, are long lived
and will typically decay far away from the primary vertex. This causes
differences in distributions related to the production points of the
muons. Such a distribution is the one in impact parameter. The impact
parameter is the distance from the beam axis the muon track passes a
plane orthogonal to the beam axis at the primary interaction point. If
the difference is large enough and the distribution not smeared too
much because of multiple Coulomb scattering it can be used to
discriminate muons from light meson decays. 

The distributions in impact parameter (without smearing due to multiple
Coulomb scattering) for the muons originating from charm and light
meson decays, respectively, are shown in Fig.\,\ref{fig:impact} for the
two sets of cuts discussed earlier. The distributions of muons from
charm are shown for both charm production mechanisms.  These
distributions have the same shape (within the resolution), this is a
reflection of properties of the charmed particle itself, that are
independent of the production mechanism. The figure shows that there is
a significant difference before Coulomb scattering is taken into
account, which hopefully can be used to discriminate the background
further.

\ffig{tb}{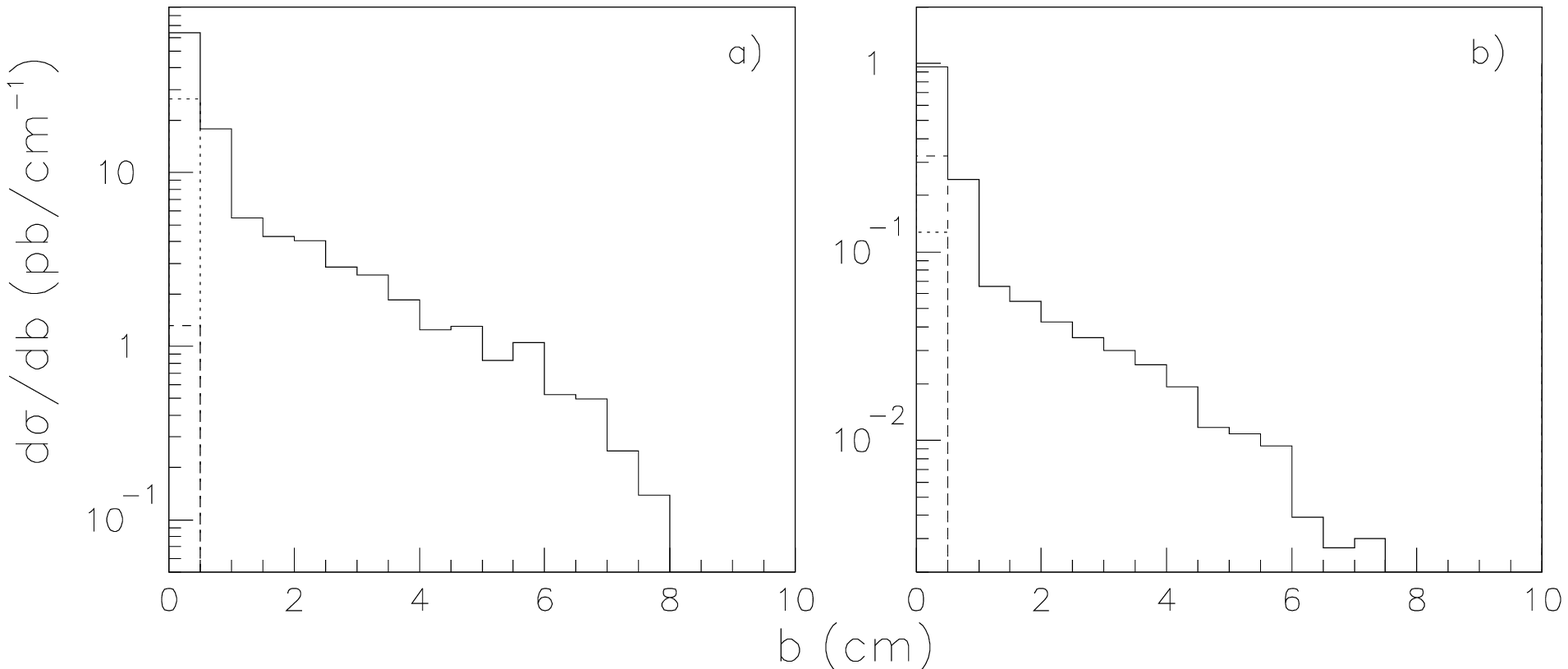}{13cm}
{\it The distribution in impact parameter for muons from light meson
decays (solid lines), intrinsic charm (dashed lines) and pQCD charm
(dotted lines). Shown are both the distribution with a) the minimal set
of cuts, and b) with the intrinsic charm search set. Decays are allowed
in the distance between Target~4 and the detector planes before the
magnet \ie $3.5\,m$.}{fig:impact}

The Coulomb scattering, which does not affect the deep inelastic
interaction rate, is important when it comes to the reconstruction of
the muon tracks. Both the scattered and the secondary muons will
undergo multiple Coulomb scattering when they pass through calorimeters
and absorbers and therefore deviate from their original direction. The
resulting scattering angle is approximately given by Gaussian
distribution having the width \cite{PDB}
\begin{equation}
\theta_0\ =\ \frac{19.2 MeV}{\beta\,c\,p}\ \sqrt{\frac{x}{X_0}}\ \left[
1\,+\,0.038\,\ln(x/X_0)\,\right],
\end{equation}
where $x/X_0$ is the thickness of the scattering material in radiation
lengths. 

The material that mainly causes the multiple
Coulomb scattering is the concrete calorimeter and the concrete
absorber. The main material in these two elements is concrete and the
radiation length for concrete ($26.7\,g/cm^2$ \cite{PDB}) is therefore
used for both elements.

Muons from Target~4 must pass through $270\,cm$ calorimeter and
absorber material (neglecting target material), mainly consisting of
concrete. This gives a width in the angular distribution of
$0.008\,rad$ for the lowest energy muons that are detectable
($10\,GeV$). The distance between Target~4 and the first layer of the
muon detector is $\sim3.5\,m$. The muons coming from charm decays will
therefore have a Gaussian distribution in impact parameter of width
$\sim3\,cm$, instead of the distribution shown in
Fig.\,\ref{fig:impact}. This means that only muons with impact
parameters larger than $\sim3\,cm$ can be excluded, which is not an
efficient cut to reduce the flux of muons from light meson decays as
can be seen in Fig.\,\ref{fig:impact}. Since the width is decreasing
with energy a tighter cut could be used if only higher energy muons
where used. This is however not preferable since the rate would then be
significantly decreased as can be seen from Fig.\,\ref{fig:theta}.  An
energy dependent cut in impact parameter would circumvent this problem
and might be a working concept. For the more upstream targets the
broadening will be even larger due to the larger distance between the
target and the calorimeter/absorber. For Target~1 one must also include
scattering in the iron calorimeter, which will further enlarge the
broadening. So an energy and target dependent cut in the impact
parameter would be the only feasible solution. This has to be
further investigated with a dedicated Monte Carlo study using {\sc
Geant} \cite{Geant}, which is beyond the scope of this paper.

\section{Summary} 

In order to search for heavy quarks by looking for additional muons in
experiments like NMC the background from decays of light mesons must be
well understood. In order to make an estimate of this muon background a
simple analytic technique has been developed. The technique has been
applied to the heavy target set-up used by NMC. It was found that the
muon flux from light meson decays is reduced to a few per cent of what
it would have been without absorption. Taken all four targets together
it would be 6\%, but would be lower if only taget~4 is considered. The
possibility to use absorbers and calorimeters positioned in the beam as
targets was also analysed. It was found that the calorimeter positioned
before the absorber can be used as a `target'. The number of muons from
decays of light mesons produced in interactions in the absorber was
found to be a few times higher than those from the real targets, and
the absorber is therefore not useable as a target. The acceptance of
events in the absorber is anyhow not sufficient for using it as a
target.

A further reduction of the background from decays of light mesons might
be possible if vertex identification can be done with sufficient
resolution. Such a reduction could be based on rejecting muons whose
tracks do not point back to the primary vertex, which could be done by
a cut in impact parameter. The simple analysis performed here shows
that this method is not efficient because of multiple Coulomb
scattering of the muons.

\section*{Acknowledgments}

I am grateful to  Andreas Mucklich for providing details on the NMC
experiment. I would also like to thank Allan Arvidsson and Gunnar
Ingelman for useful discussions and for critical reading of the
manuscript.


\begin{thebibliography}{99}

\bibitem{Brodsky80} S.J.~Brodsky, P.~Hoyer, C.~Peterson and N.~Sakai,
Phys.\ Lett.\ {\bf B93} (1980) 451.\\ S.J.~Brodsky, C.~Peterson and
N.~Sakai, Phys.\ Rev.\ {\bf D23} (1981) 2745.

\bibitem{IC-emc} J.J.~Aubert \etal, Phys.\ Lett.\ {\bf 110B} (1982) 73;
Nucl. Phys. {\bf B213} (1983) 31.

\bibitem{Hoffman} E.~Hoffman and R.~Moore, Z.\ Phys. {\bf C20} (1983)
71.

\bibitem{Harris} B.W.~Harris, J.~Smith and R.~Vogt, Nucl.\ Phys.\ {\bf
B461} (1996) 181.

\bibitem{emcdimuon} J.J.~Aubert \etal, Phys.\ Lett.\ {\bf 94B} (1980)
96; Phys.\ Lett.\ {\bf 94B} (1980) 101.

\bibitem{PDB} Particle Data Group, Phys.\ Rev.\ {\bf D50} (1994) 1.

\bibitem{Lepto} G.~Ingelman, J.~Rathsman and A.~Edin,  {\sc Lepto}\,6.4
--  A Monte Carlo for Deep Inelastic Lepton-Nucleon Scattering,
DESY preprint, to appear.

\bibitem{Lund} B.~Andersson, G.~Gustafson, G.~Ingelman and
T.~Sj\"{o}strand, Phys. Rep. {\bf 97} (1983) 33.

\bibitem{Aroma} G.~Ingelman, J.~Rathsman and G.A.~Schuler,
{\sc Aroma}\,2.2 --  A Monte Carlo Generator for Heavy Flavour
Events in $ep$ Collisions,
DESY preprint, to appear.

\bibitem{Geant} {\it {\sc Geant}-Detector description and Simulation
Tool}, CERN preprint W5013, 1993. 

\end{thebibliography}
\end{document}